# Observation of the beam-size effect at HERA

Krzysztof Piotrzkowski
Institute of Nuclear Physics, Cracow*

**Abstract**

A precise measurement of the spectrum of the photons from *ep* bremsstrahlung with the ZEUS luminosity monitor at HERA is reported. The measurement shows a reduced rate compared to the Bethe-Heitler spectrum for photon energies below 5 GeV. This suppression, called the beam-size effect, is explained by the finite transverse size of the beam overlap relative to the typical impact parameter in the process of *ep* bremsstrahlung at HERA energies.

*E-email: piotrzkowski@desy.de.

# 1 Introduction

The suppression of bremsstrahlung in beam-beam collisions due to finite transverse beam sizes, called the beam-size effect in this paper, was first observed in 1982 at the $e^+e^-$ collider VEPP-4 in Novosibirsk [1]. Soon after the discovery an explanation of this effect was proposed [2]. In the following years several papers were published where different methods of calculation were used and also some predictions for future colliders were given [3]. Recently, an extensive review of these results [4] as well as some considerations in the context of the HERA collider experiments [5, 6] were published. In this article we report the measurement of the beam sizes effect at the HERA electron-proton collider.

The beam-size effect is analogous to the phenomenon of the suppression of the electron-nucleus ($eN$) bremsstrahlung due to the screening of the nucleus' charge by the atomic electrons (see, for example [7]) when momentum transfers are small and correspond to distances larger than the atom size. The connection between these two effects can be better seen when one considers the bremsstrahlung cross-sections in impact parameter space rather than in momentum transfer space. The typical impact parameter, $\rho$, depends on the momenta of the colliding particles and the energy of the radiated photon - if it is larger than the atom size for $eN$ bremsstrahlung we deal with the screening of atomic nucleus or, if it is greater than the transverse size of the interaction region (being a convolution of both beam shapes) we deal with the beam-size effect. In the latter case, however, quantum interference has to be also taken into account [4]. Both effects result in a decrease of the bremsstrahlung cross-sections compared to the situation of infinite atom and beam sizes.

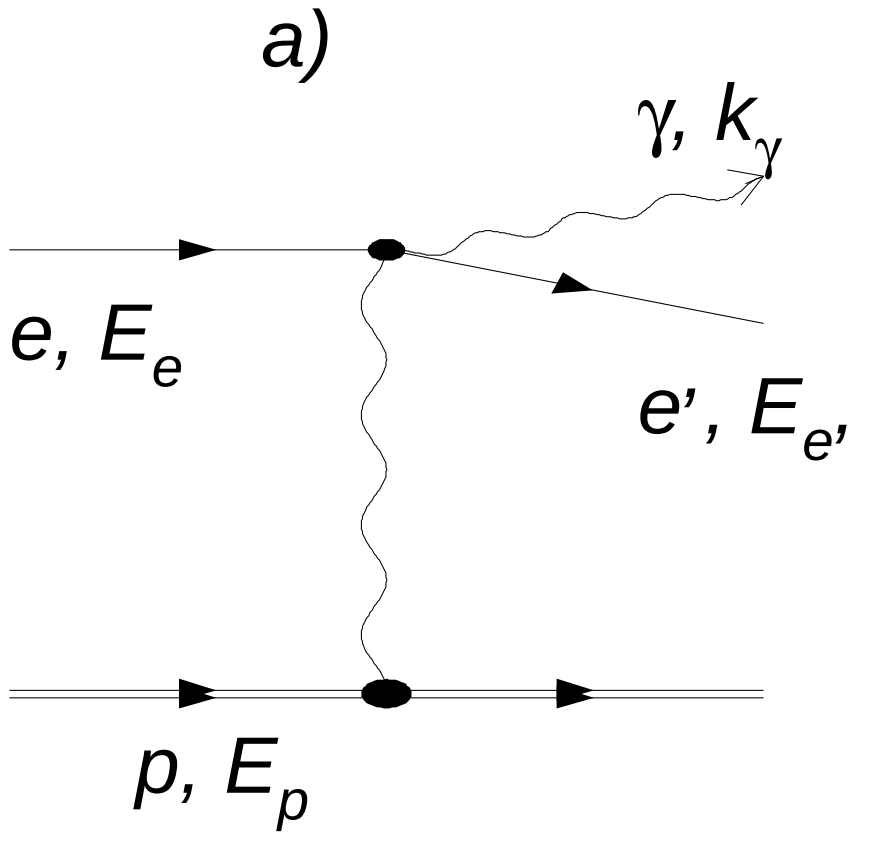

Figure 1: Feynman diagrams of $ep \to e'\gamma p$ (a) and $eN \to e'\gamma N$ (b) bremsstrahlung processes.

HERA - the high energy electron-proton ($ep$) collider based at DESY, Hamburg, provides a good opportunity to verify the theoretical predictions for the beam-size effect. For the HERA electron beam energy, $E_e$, of 27.5 GeV and proton beam energy, $E_p$, of 820 GeV the typical impact parameter in $ep$ bremsstrahlung (see [4]), $\rho \leq 4(E_e - k_\gamma)E_eE_p/m_e^2m_pk_\gamma$ (where $m_e, m_p$ are electron and proton masses, respectively, and $k_\gamma$ is the bremsstrahlung photon energy), can reach macroscopic values. For example, if $k_\gamma$=1 GeV then $\rho \leq 2\ mm$ which is bigger than the lateral sizes of the HERA beams. Therefore, one expects a significant suppression, compared to the calculation with infinite transverse beam size, of the bremsstrahlung yield at HERA for photon energies of a few GeV and less.

In 1994 HERA operated in electron-proton and positron-proton modes but since all considerations reported in this article are equally valid for electrons and positrons we will not distinguish between them, referring to electrons in all cases. There were 153 bunches stored in each ring which collided 'head-on' in two interaction regions, south and north, where the ZEUS and the H1 experiments are located, respectively. Additionally, 15 electron and 18 proton non-colliding 'pilot' bunches were stored for background studies.

In 1985, the measurement of the $ep$ bremsstrahlung rate was proposed as a way of monitoring of the HERA luminosity [8]. The ZEUS and H1 collaborations adopted this technique for luminosity measurement and built detectors to measure small angle photons and electrons. The analysis presented in this paper is based on the data taken with the ZEUS luminosity detector.

# 2 Luminosity Monitor of the ZEUS Experiment

The detailed description of the ZEUS luminosity detector can be found elsewhere [9]. Here only the 'branch' used for the bremsstrahlung photon detection is briefly described. The photon detector is located about 100 m from the interaction point (IP), upstream of the proton beam (see Fig. 2). It consisted of a 1.5 mm thick circular (100 mm in diameter) copper-beryllium exit window in the HERA high-vacuum chamber at 92 m from the IP, a 12 m long vacuum pipe, a $(1.5+2)X_0$ carbon-lead absorber ($1X_0 \equiv$ one radiation length) shielding against synchrotron radiation, and a 22 $X_0$ deep, lead-scintillator 'sandwich' calorimeter at about 107 m from the IP. The calorimeter sampling step was 1 $X_0$ with a first scintillator plate 4 times thicker than the nominal 0.26 mm thickness, to compensate for the energy loss in the absorber (3.7 $X_0$ in total). The signal from the calorimeter scintillator plates was read out by two (up and down) wavelength shifter plates connected by light guides to two photomultiplier tubes. The calorimeter had a position detector with separate readout electronics at a depth of $3X_0$.

Two fast XP2081 (Phillips) photomultiplier tubes produced signals shorter than the 96 ns time interval between HERA beam buckets. The signals were digitized in a 10.4 MHz charge-sensitive 8-bit FADC board (located close to the detector) and the digital data was transferred at 10.4 MHz over 200 m to the VME-bus based data acquisition system. The data used in this analysis is a sample taken for monitoring and calibration purposes during the 1994 ZEUS running period.

The combined system, the calorimeter and the filter, was tested at the DESY 1–5 GeV electron beam. The measurements showed that in this energy range deviations from a linear response were below 1% and that the relative energy resolution, $\sigma_E/E$, was about $0.25/\sqrt{E(GeV)}$. The filter deteriorated the calorimeter energy resolution and absorbed an approximately constant amount of energy for electrons above 1 GeV. The non-uniformity of the calorimeter response was below 1% in the fiducial volume of the bremsstrahlung events. The non-linearity of the readout electronics and FADC boards was determined with the use of LED (light-emitting-diode) light pulse measurements. It was well described as a quadratic deviation from linear behaviour and found to be 0.07% deviation per GeV.

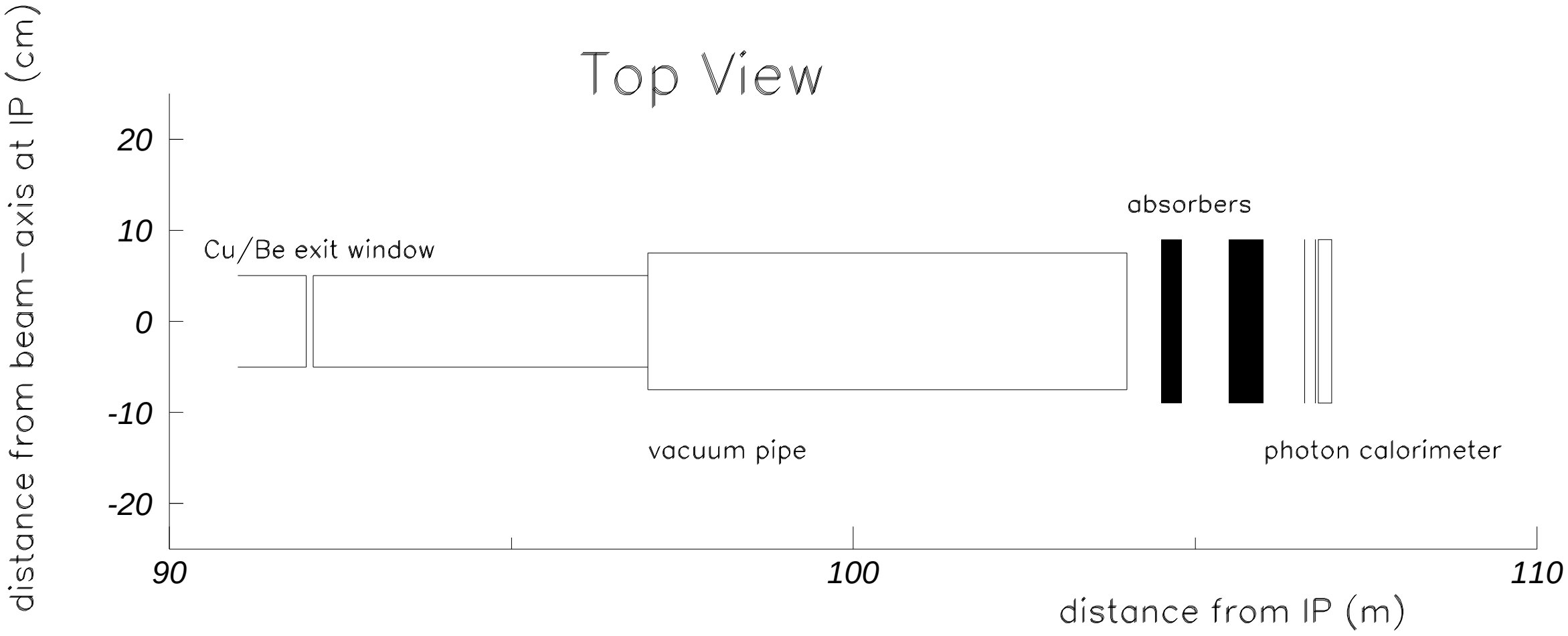

Figure 2: Layout of the photon detector in the ZEUS luminosity monitor.

The geometrical acceptance of the photon detector for $ep$ bremsstrahlung was more than 99% and was sensitive only to the angular dispersion and the direction (tilt) of the electron beam at the IP. The width of the angular dispersion was typically 0.15 mrad horizontally, and 0.05 mrad vertically. Both, the dispersion and the tilt of the electron beam were not significantly changing during the running period and were controlled by the measurement of the bremsstrahlung photon beam profile with the position detector. The majority of the events was well contained in the detector whose transverse dimensions ($18 \times 18\ cm^2$) were much larger than the lateral size of the photon beam at 107 m from the IP ($\approx 6 \times 2\ cm^2$).

Typical event rates during the 1994 running period were about $3 \times 10^5$ events/s with at least 1 GeV deposited in the photon detector. Therefore, the probability of more than one bremsstrahlung event in one bunch crossing was sizable, e.g. a few percent of events with more than 0.1 GeV deposit was actually due to pileup of two bremsstrahlung events resulting from the same beam collision.

# 3 Measurement of the $eN$ Bremsstrahlung Spectrum

For the analysis presented in this article about $10^7$ events of $ep$ and $10^6$ events of $eN$ bremsstrahlung were used. The events were collected by random sampling of the data flow in the data acquisition system of the ZEUS luminosity monitor, retaining those events which deposited more than 1 GeV in the photon calorimeter. The data was taken during six runs lasting between 6 and 11 hours. Bremsstrahlung from interactions of the electrons from the pilot bunches with the nuclei of the rest gas molecules in the ZEUS beam pipe was measured in order to estimate the contribution of $eN$ bremsstrahlung for the colliding bunches and to check the understanding of the energy scale in the photon measurement. The background from the proton beam halo was negligible while the signal due to synchrotron radiation which was typically less than 0.02 GeV was controlled by monitoring the ADC pedestals.

As mentioned, the spectra of the $eN$ bremsstrahlung measured with the pilot bunches were used to check the photon energy measurement. This was done by fitting the data with the function, $F[E_\gamma(ADC_\gamma)]$, being a convolution of the $eN$ bremsstrahlung differential cross-section [10], $d\sigma_{eN}/dk_\gamma$, and a Gaussian function describing the calorimeter energy resolution:

$$F[E_\gamma(ADC_\gamma)] = \frac{P1}{\sqrt{2\pi}}|J| \int_{0.1GeV}^{E_e - m_e} \frac{\mathrm{d}\sigma_{eN}}{\mathrm{d}k_\gamma} \exp\left[-(E_\gamma - k_\gamma)^2/2\Delta_{det}^2\right] \frac{\mathrm{d}k_\gamma}{\Delta_{det}}\,, \tag{1}$$

where:

$$\frac{\mathrm{d}\sigma_{eN}}{\mathrm{d}k_\gamma} = 4\alpha r_e^2 \frac{E_{e'}}{k_\gamma E_e}\left[\left(\frac{E_e}{E_{e'}} + \frac{E_{e'}}{E_e} - \frac{2}{3}\right)\left[Z^2 \ln\left(184.15 Z^{-1/3}\right) + Z \ln\left(1194 Z^{-2/3}\right)\right] + \frac{1}{9}(Z^2 + Z)\right],$$

$E_\gamma$ is the measured photon energy calculated from the equation,
$ADC_\gamma = P2(E_\gamma - P3)[1 + f_{nl}(E_e - E_\gamma)] + ADC_0$, where $ADC_\gamma$ is the average signal of the two readout channels and $ADC_0$ is the measured average pedestal,
$|J|$ is the Jacobian of the $E_\gamma \rightarrow ADC_\gamma$ transformation,
$f_{nl}$ is the non-linearity parameter,
$E_{e'} = E_e - k_\gamma$ is the energy of the secondary electron,
$\Delta_{det} = P4\sqrt{k_\gamma}$ is the energy resolution of the calorimeter,
$\alpha$ is the fine structure constant,
$r_e$ is the classical electron radius,
$Z$ is the atomic number of the nucleus, $N$, and
P1 (normalization), P2 (calibration constant), P3 (the energy shift due to the filter), P4 (calorimeter resolution) are the fitted parameters.

The fits were performed for $ADC_\gamma > 33$ (corresponding to $E_\gamma >$3.5 GeV) assuming a mean atomic number of the nuclei[1] equal $Z = 4.2$, according to the estimated rest gas composition [11]. The quality of the fits was good - the typical $\chi^2/n.d.f.$ was 1.1, see Fig. 3. As expected, the calorimeter resolution and the energy shift due to the absorber obtained from the fit were constant from run to run (i.e. $P3 \approx 0.08$ GeV and $P4 \approx 0.265$) while the calibration constant was slightly changing due to slow variations of the photomultiplier gains.

The extrapolation of the curves obtained from the fits to lower photon energies shows an excess of events for photon energies below 2 GeV due to the Compton scattering of the blackbody photons in the beam pipe off the beam electrons. This effect was already measured at HERA using the HERA polarimeter [12]. Since the cross-section for this process and the length of the HERA straight section from which scattered photons are accepted are well-known and the current of the pilot bunches is precisely measured with the HERA pick-up coils, the absolute contribution of this process was predicted, with only one free parameter - the beam-pipe temperature, $T_{bp}$. Adding the contribution from this process to the *egas* bremsstrahlung spectrum for $T_{bp} \approx 310$ K yielded a good description of the data down to 1 GeV photon energy as shown in Fig. 3 (dashed curves). This shows that only for $E_\gamma < 2.5$ GeV the Compton scattering of the blackbody radiation contributed more than 1% to the photon spectra measured with electron pilot bunches.

[1]The fit parameters were not sensitive to the nominal value of $Z$ for a reasonable range of $Z$ values.

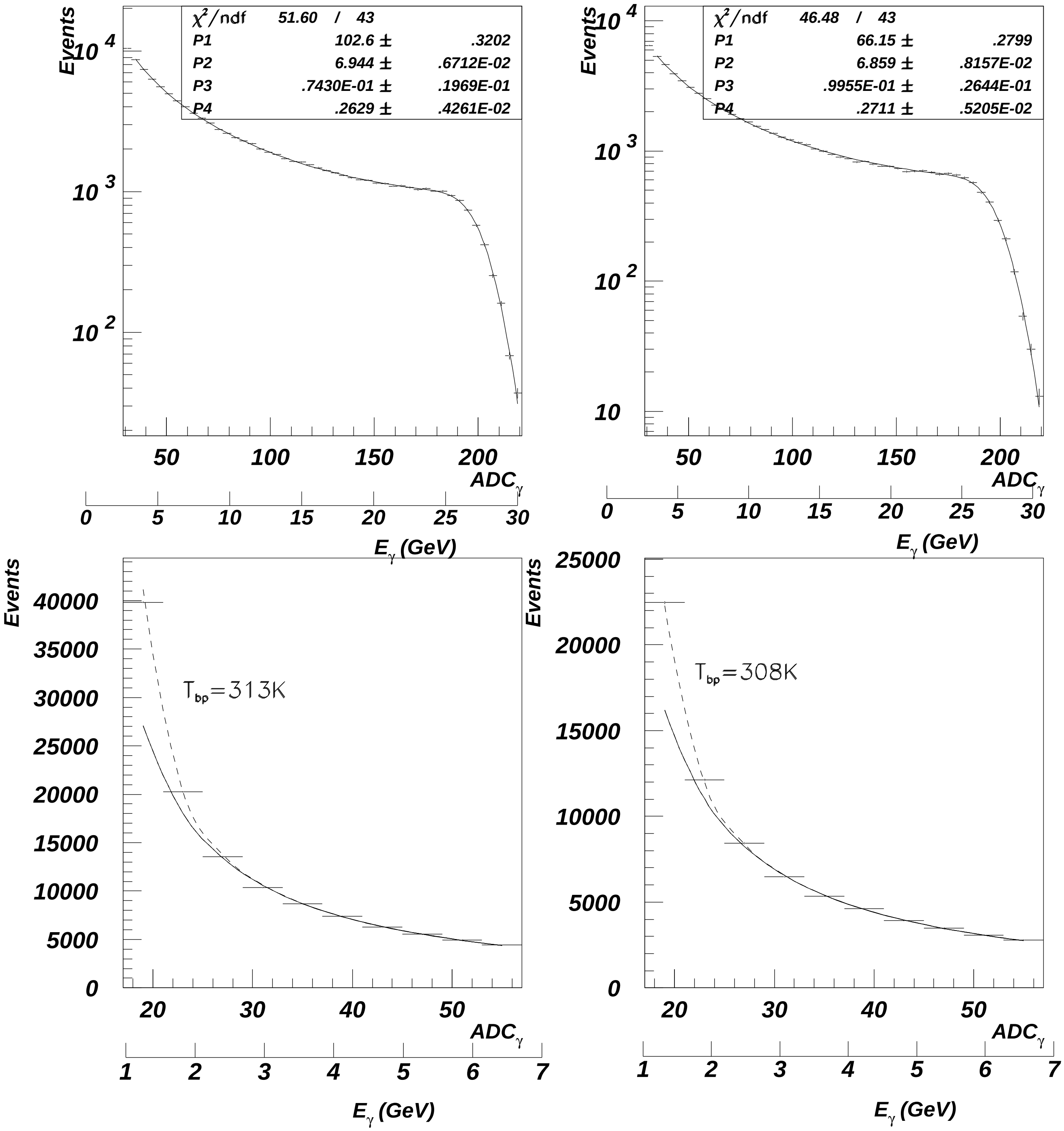

Figure 3: Two spectra of $eN$ bremsstrahlung measured in the luminosity monitor using the electron pilot bunches. The histograms represent the data and the curves are results of fitting the function $F$ (from Eq.1) for $E_\gamma > 3.5$ GeV; in the lower plots the low energy parts of the spectra are shown with extrapolations of the curves obtained from the fits - the excess of events with $2 > E_\gamma > 1$ GeV is well described by adding a contribution from Compton scattering of the blackbody photons off the beam electrons (dashed curves, $T_{bp}$ is the beam-pipe temperature).

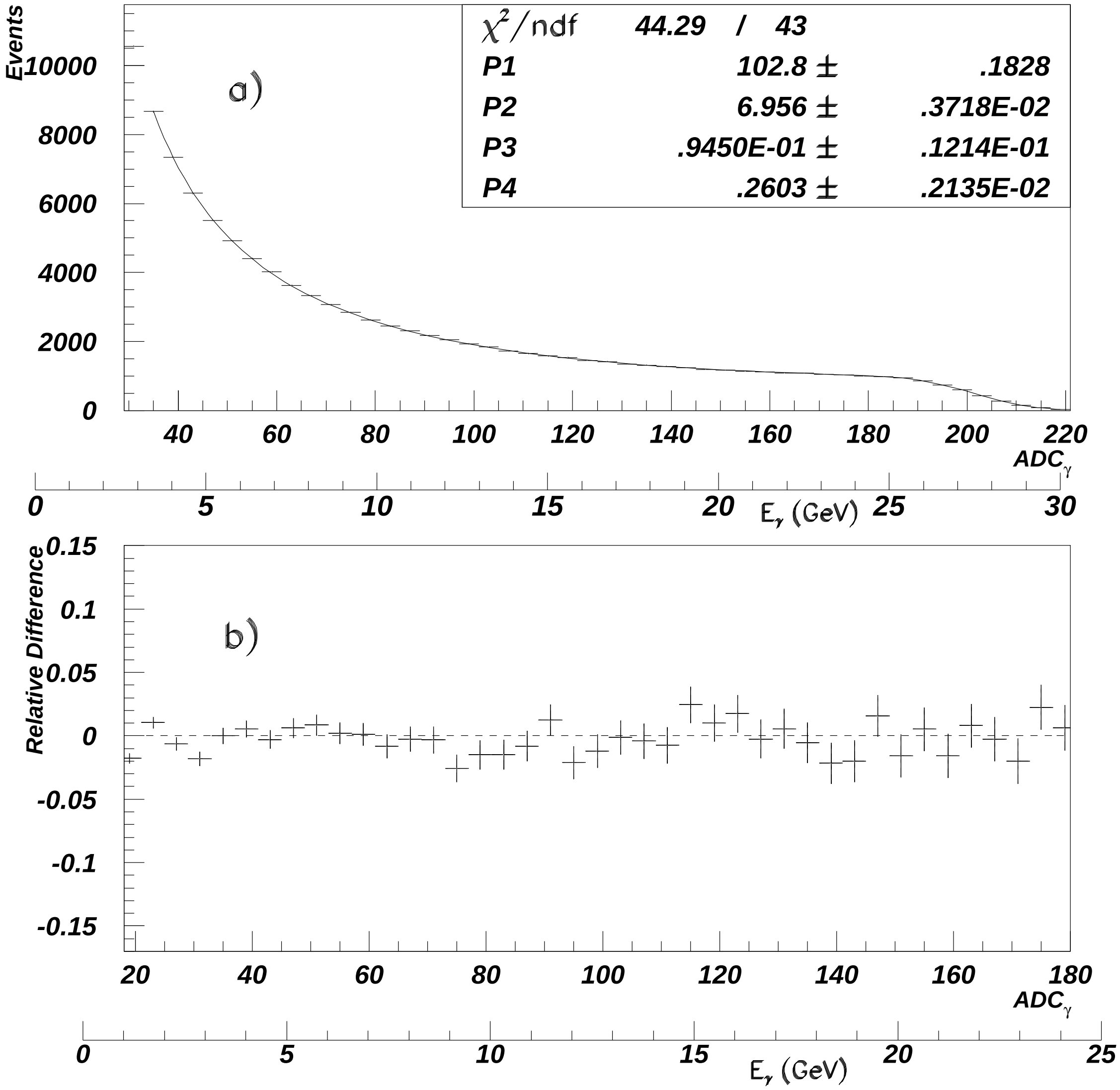

Figure 4: (a) Fit to the simulated spectrum of the $eN$ bremsstrahlung; (b) relative difference between measured and simulated spectra of the $eN$ bremsstrahlung.

# 4 Monte Carlo Simulation

Based on the measurements of the $eN$ bremsstrahlung a simple and fast Monte Carlo simulation of the bremsstrahlung measurement was written. It included: generation of $ep$ and $eN$ bremsstrahlung events with the BREMGE program [6]; generation of events when a thermal photon was scattered in the Compton process; simulation of the event pileup when two or more events occured during a single beam collision by generation of the event multiplicity from the Poisson distribution; the energy shift of 0.135 GeV due to the absorber[2] and the energy smearing due to the calorimeter resolution; the non-linearity of the readout electronics and the digitization of the calorimeter signal (i.e. the conversion from energy to ADC counts, including the noise of the electronics).

[2]It was found from the MC simulation that fits underestimate the shift due to the filter by about 0.05 GeV, see Fig. 4a were the MC spectrum generated with a 0.135 GeV shift was fitted.

This allowed the photon energy spectra to be simulated separately for each data sample (taken in different running conditions) with high statistics (more than $10^8$ events were generated). In Fig. 4b the measured distribution is compared with the result of the MC simulation showing a very good agreement; the MC spectrum was normalized to the data by requiring the same number of events for $E_\gamma > 12$ GeV.

For the simulation of the bremsstrahlung measurement with the colliding bunches two more parameters, beside the parameters determined from the $eN$ data, were required - the relative contribution of $ep$ and $eN$ events and the average multiplicity of the bremsstrahlung events in a single beam collision, $\mu_{br}$, needed for simulation of the event pileup. The former was obtained by scaling the rate of events (for $16 > E_\gamma > 10$ GeV) measured with the electron pilot bunches with the ratio of currents of colliding and pilot electron bunches and comparing it with the total event rate for the colliding bunches. In runs used for this analysis the $eN$ bremsstrahlung contribution was 5–8%. The latter was determined from the ratio of the total event rate for colliding bunches to the rate of the beam collisions, $f_{coll}$ (for runs considered here, $f_{coll} = 153 \times 47.3$ kHz=7.24 MHz). One should note that the simulation of the event pileup was essential as it also depleted the low-energy part of the $ep$ bremsstrahlung spectra and could mimic the beam-size effect. Typically, the distortion of the bremsstrahlung spectrum due to event pileup was comparable to the beam-size effect. The simulation of event pileup was however well controlled by the comparison of the number of well identified 'multiple' events in data and MC, when the event energy exceeded substantially the electron beam energy.

# 5 Measurement of the Beam-Size Effect

In Fig. 5 the comparison is shown between the MC prediction for the bremsstrahlung spectra for the colliding bunches. The MC sample was normalized to the data in the same way as for the bremsstrahlung measured with the pilot bunches. It can be seen that if the Bethe-Heitler formula [13] is used for the generation of events, the MC points overshoot the data for photon energies below 8 GeV and the difference reaches about 5% for 2 GeV photons. The difference however disappears once the beam-size effect is taken into account in the MC simulation. The simulation also shows that the measured relative difference (see Fig. 5c) describes well the expected difference between the Bethe-Heitler cross-section and the modified cross-section. Therefore, one can directly compare such distributions with the theoretical predictions (open dots).

All relevant formulae for calculation of the beam-size effect in $ep$ bremsstrahlung at HERA and used in this analysis can be found in [4]. In calculations (and in the MC simulation) we assumed that the HERA beams had Gaussian shapes and collided 'head-on'. The shape of the HERA proton beam was routinely measured by wire-scans and the shape of the interaction region, being the convolution of the shapes of two beams, was measured twice by scanning of the electron beam with the proton beam. No sigificant deviation from the Gaussian shape was observed. The nominal transverse width of the interaction region, $\Delta_h = 350$ $\mu m$ horizontally, and $\Delta_v = 70\ \mu m$ vertically were determined from the scans[3]. In Tab. 1 the corrections are listed for a $8 > k_\gamma > 2$ GeV photon energy interval and for non-nominal width of the interaction region as well as for non-central beam collisions for which the beam impact parameters, $d_h, d_v$, are non-zero.

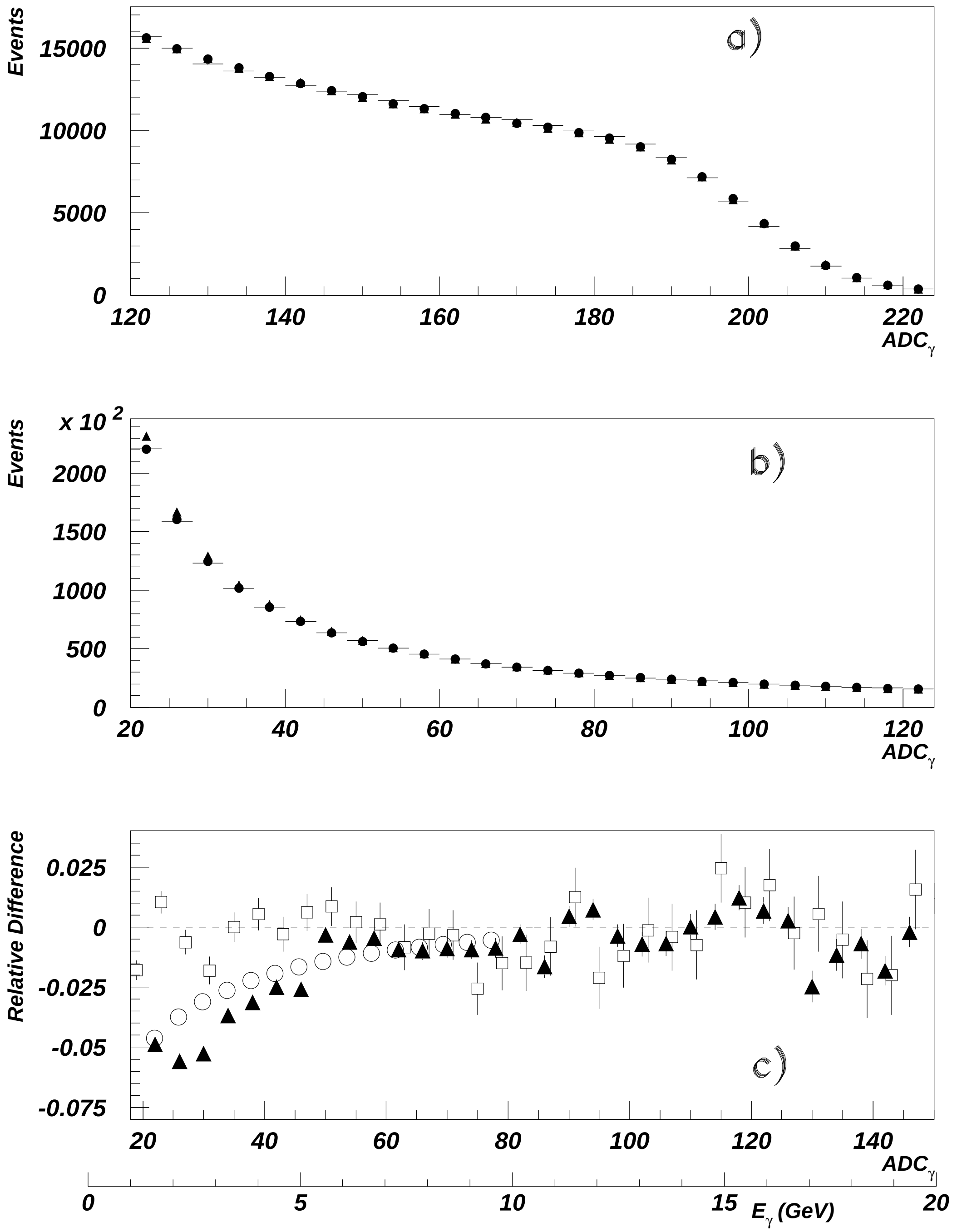

Figure 5: Comparison of high (a) and low (b) energy parts of the measured (histogram) and the simulated spectra of $ep$ bremsstrahlung with (dots) and without (triangles) beam-size effect; (c) the relative differences - triangles: between the measured and simulated (without beam-size effect) $ep$ bremsstrahlung spectra, squares: between the measured and simulated $eN$ bremsstrahlung spectra, open dots: between the cross-section corrected for the beam-size effect and the standard Bethe-Heitler cross-section (small systematic fluctuations of the measured points were due to unequal widths of the ADC bins).

Table 1: Relative Corrections, $\sigma_{corr}$, to the Bethe-Heitler Cross-section

| Beam parameters | $\sigma_{corr}(\%)$ |
|---|---|
| nominal $\Delta_h, \Delta_v, d_h, d_v$ | -3.28 |
| nominal $\Delta_h, d_h, d_v; \Delta_v = 77\mu m$ | -3.06 |
| nominal $\Delta_v, d_h, d_v; \Delta_h = 385\mu m$ | -3.26 |
| nominal $\Delta_h, \Delta_v, d_h; d_v = 20\mu m$ | -3.57 |
| nominal $\Delta_h, \Delta_v, d_v; d_h = 100\mu m$ | -3.57 |

Table 2: Run Parameters

| Parameter | Run#1 | Run#2 | Run#3 | Run#4 | Run#5 | Run#6 |
|---|---|---|---|---|---|---|
| Rate($16 > E_\gamma > 10$ GeV) (kHz) | 45 | 43 | 18.5 | 29 | 37.5 | 44 |
| $eN \rightarrow e\gamma N$ contribution | 0.08 | 0.065 | 0.055 | 0.061 | 0.054 | 0.057 |
| $\mathcal{L}_{spec}$ $(10^7 cm^{-2}s^{-1})$ | 2.7 | 2.4 | 1.9 | 1.9 | 2.5 | 2.7 |

To quantify better the measured deviation from the Bethe-Heitler spectrum we introduced the relative difference, $r_{diff}$, between the number of events (after subtraction of the contribution of the $eN$ bremsstrahlung) in the $8 > E_\gamma > 2$ GeV range in the data, $N_{data}^{8>E_\gamma>2GeV}$, and in the MC sample without the beam-size effect, $N_{MC}^{8>E_\gamma>2GeV}$:

$$r_{diff} = 1 - \frac{N_{MC}^{8>E_\gamma>2GeV}}{N_{data}^{8>E_\gamma>2GeV}}. \tag{2}$$

This quantity was determined from the data samples collected in quite different running conditions, see Tab. 2.

Since $r_{diff}$ can be directly compared with theoretical predictions, in Fig. 6a the measured $r_{diff}$ is shown with the theoretical expectations where error bars reflect both statistical and systematic error contributions, as is discussed below. In Fig. 6b the correlation between the measured and expected values of $r_{diff}$ are shown. The typical value of $r_{diff}$ measured for the $ep$ bremsstrahlung was about $-0.04$.

[3]The widths of the interaction region for centrally colliding beams are defined as follows: $\Delta_h = \sqrt{\sigma_{h,e}^2 + \sigma_{h,p}^2}$ and $\Delta_v = \sqrt{\sigma_{v,e}^2 + \sigma_{v,p}^2}$, where $\sigma_{h,e}, \sigma_{h,p}, \sigma_{v,e}, \sigma_{v,p}$ are the horizontal and vertical widths at the IP of the electron and proton beam, respectively.

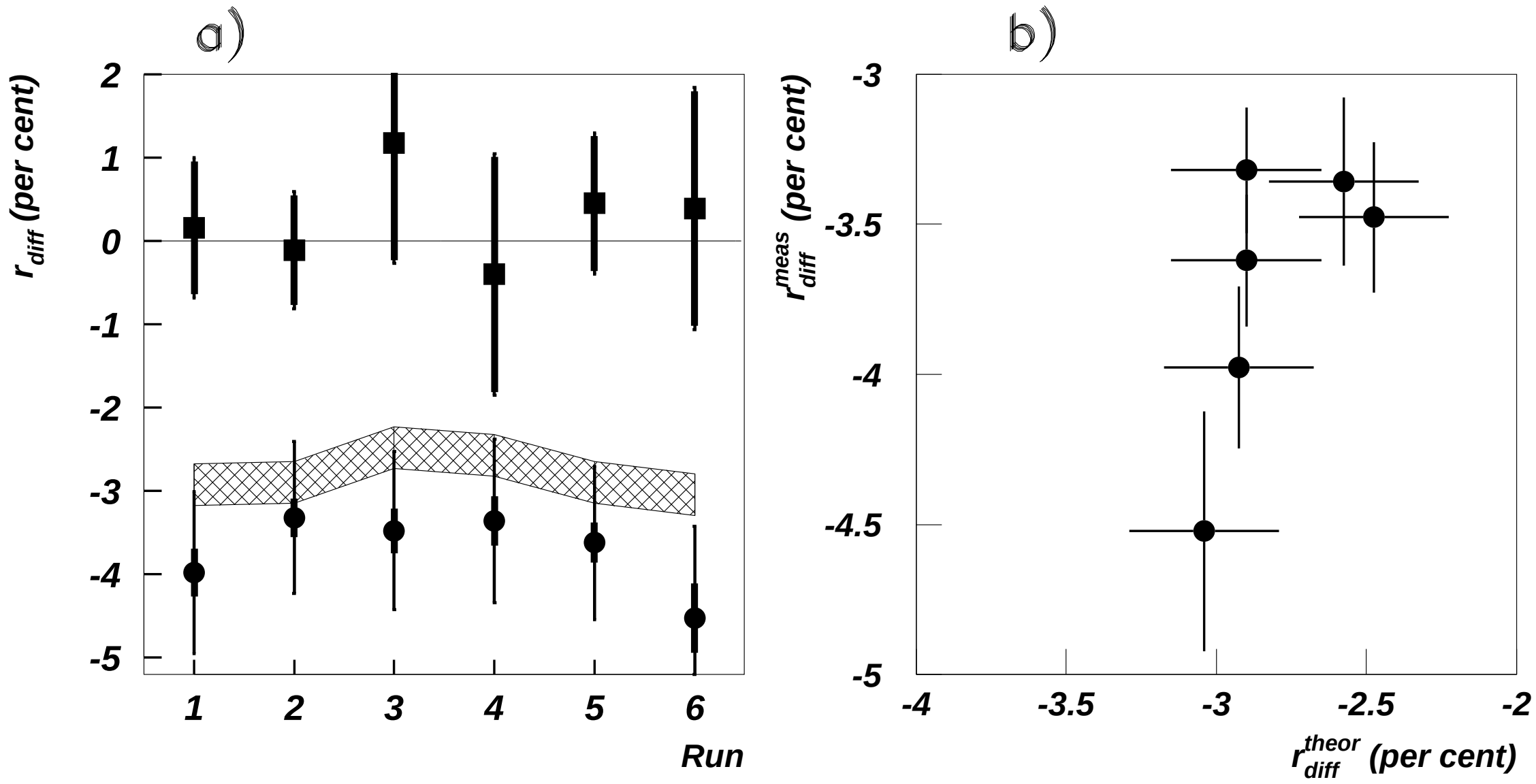

Figure 6: (a) Comparison of the measured (dots with thick and thin error bars representing statistical and total errors, respectively) and predicted (hatched area) values of $r_{diff}$, for reference the $r_{diff}$ measured for $eN$ bremsstrahlung is also shown (squares); (b) correlation between the measured and predicted values of $r_{diff}$, the vertical error bars represent statistical errors only.

# 6 Error Analysis

The steep rise of the $ep$ bremsstrahlung spectrum with the decreasing photon energy and high event rate made the measurement of the beam-size effect in at HERA a demanding and difficult task. At the end, the measurement of the beam-size effect was verified by a simultaneous measurement of the bremsstrahlung spectrum from electron-gas collisions which agreed well with the theoretical predictions, i.e. the $r_{diff}$ measured for the $eN$ bremsstrahlung was consistent with zero (see Fig. 6a).

The statistical errors were negligible in the determination of $r_{diff}$ but both experimental data and theoretical predictions had clear sources of systematic errors. The major uncertainty in the calculation of the modification of the Bethe-Heitler cross-section from [4, Eq. 6.4] was the transverse shape of the interaction region, which is given by the shape of the two beams and their separation. At HERA, the size of the interaction region, $2\pi\Delta_h\Delta_v$, was quite precisely determined from the measured specific luminosity, $\mathcal{L}_{spec}$, as $\mathcal{L}_{spec} = f_{rev}/2\pi\Delta_h\Delta_v$, where $f_{rev} = 47.317$ kHz is the revolution frequency of the HERA beams. The ratio of the horizontal to vertical widths of the interaction region, also relevant for the evaluation of the beam-size effect, see Tab. 1, was however not known so precisely. The expected ratio was 5:1 and we assumed that it to be in the range 6:1 to 4:1. The centering of the beams was being performed by maximising the specific luminosity and the typical impact parameter of the beams should not be larger than 25% of the width of the interaction region. The first uncertainty resulted in a $\pm 0.0015$ and the second in a $\pm 0.002$ error in the theoretical prediction of $r_{diff}$ summing up to an 0.0025 total theoretical uncertainty. In the determination of the experimental systematic error the uncertainty due to absorption in the filter was dominating. The corrections due to the contribution of the $eN$ bremsstrahlung and the Compton scattering of the blackbody radiation were small and lead to errors on $r_{diff}$ of less than 0.0014 and $3 \times 10^{-4}$, respectively. The uncertainty due to determination of the calorimeter resolution was negligible and a 0.15% uncertainty of the calibration constant caused a 0.002 error. The pedestal uncertainty of 0.03 ADC channel resulted in a 0.002 error, while the uncertainty in the determination of the non-linearity of the readout electronics, $f_{nl} = (7 \pm 2) \times 10^{-4}$/GeV, resulted in a $\pm 0.0013$ error on $r_{diff}$. The energy shift determined from the fits to the data and the results of the test beam measurements was constrained to $(0.135 \pm 0.015)$ GeV range resulting in a 0.006 uncertainty. The event pileup was known to 3%, resulting in a 0.001 error. The systematic errors were combined in quadrature resulting in an overall systematic error in the determination of $r_{diff}$ of 0.007. The contributions to the experimental and theoretical systematic error are summarized in Tab. 3.

Table 3: Systematic errors in the measurement and calculation of $r_{diff}$.

| Source | $\Delta(r_{diff})$ |
|---|---|
| Calibration constant | 0.002 |
| Non-linearity due to absorber | 0.006 |
| Non-linearity in electronics | 0.0013 |
| Background subtraction | 0.0014 |
| Event pile-up | 0.001 |
| Calorimeter resolution | 0.0002 |
| ADC pedestals | 0.002 |
| Total error (exper.) | 0.007 |
| Beam widths | 0.0015 |
| Beam separation | 0.002 |
| Total error (theor.) | 0.0025 |

# 7 Conclusions

The beam-size effect, i.e. the suppression of $ep$ bremsstrahlung due to finite beam sizes, has been measured at the HERA collider by observing the small deviation of the measured electron-proton bremsstrahlung spectrum from the Bethe-Heitler spectrum at low photon energies. The difference for the $ep$ bremsstrahlung between the data and the prediction based on standard cross-sections was found to be $(-3.7 \pm 0.7)\%$ for $8 > k_\gamma > 2$ GeV photon energy range. This was accounted for by the beam-size effect which was predicted to be $(-2.8 \pm 0.25)\%$ for the same bremsstrahlung energy range.

# 8 Acknowledgments

The author would like to thank very much Dr. Robert Klanner for many useful comments and suggestions. I am also very grateful to my colleagues from the ZEUS luminosity group for their diligent efforts in running the luminosity monitor and allowing me to use its data. I thank very much the DESY directorate for the support of my stay at DESY laboratory where this work was done.